# Comment on
# "Group method analysis of magneto-elastico-viscous flow along a semi-infinite flat plate with heat transfer", M.M. Helal and M. B. Abd-el-Malek, Journal of Computational and Applied Mathematics 173 (2005) 199-210.

Asterios Pantokratoras
Associate Professor of Fluid Mechanics
School of Engineering, Democritus University of Thrace,
67100 Xanthi – Greece
e-mail:apantokr@civil.duth.gr

In the above paper the authors treat the boundary layer flow of a elasto-viscous liquid along an infinite plate in the presence of a transverse magnetic field. The plate temperature is higher than the ambient fluid temperature. The boundary layer equations are transformed into ordinary ones using the group theory and subsequently are solved numerically. Velocity, temperature, shear stress and heat transfer profiles are presented for values of magnetic parameter M=0, 0.5 and 1. This is an interesting work but there are some fundamental errors which are presented below:

The governing momentum equation (equation 2.1 in their paper) is

$$\frac{\partial \psi}{\partial y}\frac{\partial^2 \psi}{\partial x \partial y} - \frac{\partial \psi}{\partial x}\frac{\partial^2 \psi}{\partial y^2} = \frac{\partial^3 \psi}{\partial y^3} - \frac{M}{x}\frac{\partial \psi}{\partial y} - k\left[\frac{\partial \psi}{\partial y}\frac{\partial^4 \psi}{\partial x \partial y^3} - \frac{\partial \psi}{\partial x}\frac{\partial^4 \psi}{\partial y^4} + \frac{\partial^2 \psi}{\partial x \partial y}\frac{\partial^3 \psi}{\partial y^3} - \frac{\partial^2 \psi}{\partial y^2}\frac{\partial^3 \psi}{\partial x \partial y^2}\right] \quad (1)$$

where M is the magnetic parameter and k is a nondimensional elastic parameter representing the non-Newtonean character of the fluid. Apparently the above equation is valid for a Newtonean fluid when k=0. In page 200 it is mentioned that "the governing equations of motion are given in [6]" where [6] is the work of Beard and Walters (1964). However equation (1) does not exist in Beard and Walters (1964). These authors presented an equation in dimensional form without magnetic field (equation 17 in their work). Apparently Helal and Abd-el-Malek (2005) transformed equation 17 by Beard and Walters (1964), using the usual definition of the stream function ψ and added the term which represents the influence of the magnetic field (the second term on the right hand side). It should be noted that no definition for M and k and $\psi$ has been given in the work of Helal and Abd-el-Malek (2005). To make things simple we assume that the fluid is Newtonean and therefore equation (1) becomes

$$\frac{\partial \psi}{\partial y}\frac{\partial^2 \psi}{\partial x \partial y} - \frac{\partial \psi}{\partial x}\frac{\partial^2 \psi}{\partial y^2} = \frac{\partial^3 \psi}{\partial y^3} - \frac{M}{x}\frac{\partial \psi}{\partial y} \tag{2}$$

The relation between velocities and the stream function $\psi$ is (Beard and Walters, 1964, equation 19)

$$u = \frac{\partial \psi}{\partial y} \tag{3}$$

$$v = -\frac{\partial \psi}{\partial x} \tag{4}$$

Using the above equations we can go back to the original momentum equation of a Newtonean fluid and equation (2) becomes

$$u\frac{\partial u}{\partial x} + v\frac{\partial u}{\partial y} = \frac{\partial^2 u}{\partial y^2} - \frac{M}{x}u \tag{5}$$

The boundary conditions are (equations 2.3 and 2.4 in their work)

at y = 0:   u = v = 0 (6)
as y → ∞   u = $U_0$ (7)

where $U_0$ is a constant (page 200). Let us apply now the momentum equation (5) at large y. At large distances from the plate the velocity is everywhere constant and equal to $U_0$ and therefore the velocity gradient $\partial u/\partial y$ is zero. This means that the momentum equation takes the following form at large y

$$U_0 \frac{\partial U_0}{\partial x} = -\frac{M}{x}U_0 \tag{8}$$

or

$$\frac{\partial U_0}{\partial x} = -\frac{M}{x} \tag{9}$$

From the above equation we see that the free stream velocity should change along x and therefore the momentum equation (5) and subsequently the equation (2) and the initial equation (1) are not compatible with the boundary condition that the free stream velocity is constant (equation 7). This is an error made frequently in the literature (see Vajravelu, 2007 and Xu, 2007). Taking into account the above argument it is clear that the momentum equation (1) is wrong and the results presented by Helal and Abd-el-Malek (2005) also wrong.

In page 206 the boundary conditions for velocity and temperature profiles are given for the transformed equations at η=0 and ∞ (equations 3.39-3.40) where η is the transverse similarity variable. Velocity becomes 1 (page 207) while temperature becomes 0 as η → ∞. It is known in boundary layer theory that velocity and temperature profiles approach the ambient fluid conditions asymptotically as η → ∞ and do not intersect the line which represents the boundary conditions. Some velocity and temperature profiles that approach the ambient conditions correctly (asymptotically) in a boundary layer flow are shown in Arpaci and Larsen (1984, page 154), in Cebeci and Bradshaw (1988, page 42), in Kakac and Yener (1995, page 47), in Bejan (1995, page 43), in Incropera and DeWitt (1996, page 290), in Oosthuizen and Naylor (1999, page 62), in Schlichting and Gersten (2003, pages 215, 265 and 281) and in White (2006, page 80). In figure 1 of the present work we show a velocity profile taken from figure 1(b) of the above work. We see that the velocity profile does not approach the ambient velocity asymptotically but intersects the horizontal line ($F_0^{'}$=1) almost vertically. At the same figure we show the correct shape of this velocity profile. It is clear that this velocity profile given by Helal and Abd-el-Malek (2005) is wrong and the same happens with the velocity profile for M=0.5 included in figure 1(b) and the two temperature profiles for M=0.5 and M=1 in figure 2(a). It should be noted here that these two temperature profiles are almost straight lines and in boundary layer flow straight line velocity or temperature profiles do not exist. The only velocity and temperature profiles which are correct are those for M=0. It is clear that the profiles which do not approach the ambient conditions asymptotically are truncated due to a small calculation domain used. The authors used for all the above cases a calculation domain with $\eta_{max}=6$. However this calculation domain was not sufficient to capture the real shape of profiles and a wider calculation domain, greater than 6, should be used. It is sure that the truncation of these profiles has introduced errors in the corresponding shear stress profiles (M=0.5 and M=1 in figure 1(c)) and heat transfer profiles (M=0.5 and M=1 in figure 2 (b)).

In conclusion there are two fundamental errors in the above work and the presented results are inaccurate.

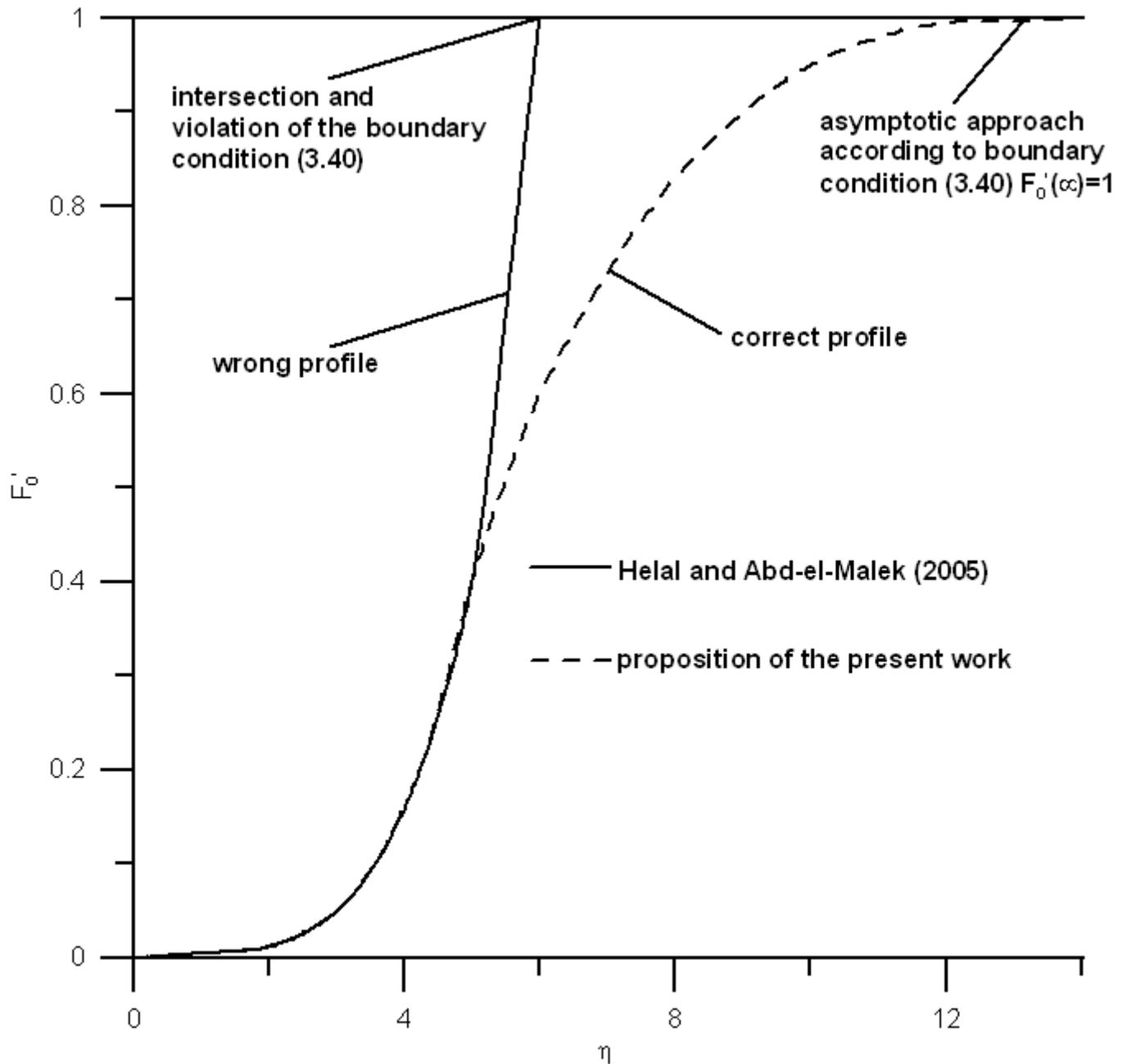

Figure 1. The solid line represents a dimensionless velocity profile for M=1 and Pr=0.7. This profile has been reproduced from figure 1(b) by Helal and Abd-el-Malek (2005). The dashed line profile is in agreement with the boundary layer theory.